\documentclass[sigconf]{acmart} 

\AtBeginDocument{%
  \providecommand\BibTeX{{%
    \normalfont B\kern-0.5em{\scshape i\kern-0.25em b}\kern-0.8em\TeX}}}

\setcopyright{acmcopyright}
\copyrightyear{2024}
\acmYear{2024}
\setcopyright{rightsretained}
\acmConference[CHI '24]{Proceedings of the 2024 CHI Conference on Human Factors in Computing Systems}{May 11--16, 2024}{Honolulu, Hawai'i}
\acmBooktitle{Proceedings of the 2024 CHI Conference on Human Factors in Computing Systems (CHI '24), May 11--16, 2024, Honolulu, Hawai'i}\acmDOI{10.1145/3544548.3581357}
\acmISBN{978-1-4503-9421-5/23/04}

\graphicspath{{Graphics/}}

\usepackage{multirow}
\usepackage{subcaption}
\usepackage{enumitem}
\usepackage{float}
\usepackage{color, xcolor,colortbl}
\usepackage{soul}

\begin{document}


\title[Designing for Harm Reduction]{Designing for Harm Reduction: \\Communication Repair for Multicultural Users' Voice Interactions}








\author{Kimi V. Wenzel}
\affiliation{{%
  \institution{Carnegie Mellon University}
  \streetaddress{5000 Forbes Avenue}
  \city{Pittsburgh}
  \state{Pennsylvania}
  \country{USA}
  \postcode{15213}
}}
\email{kwenzel@cs.cmu.edu}

\author{Geoff Kaufman}
\affiliation{{%
  \institution{Carnegie Mellon University}
  \streetaddress{5000 Forbes Avenue}
  \city{Pittsburgh}
  \state{Pennsylvania}
  \country{USA}
  \postcode{15213}
}}
\email{gfk@cs.cmu.edu}

\begin{teaserfigure}
\centering
  \includegraphics[width=\textwidth]{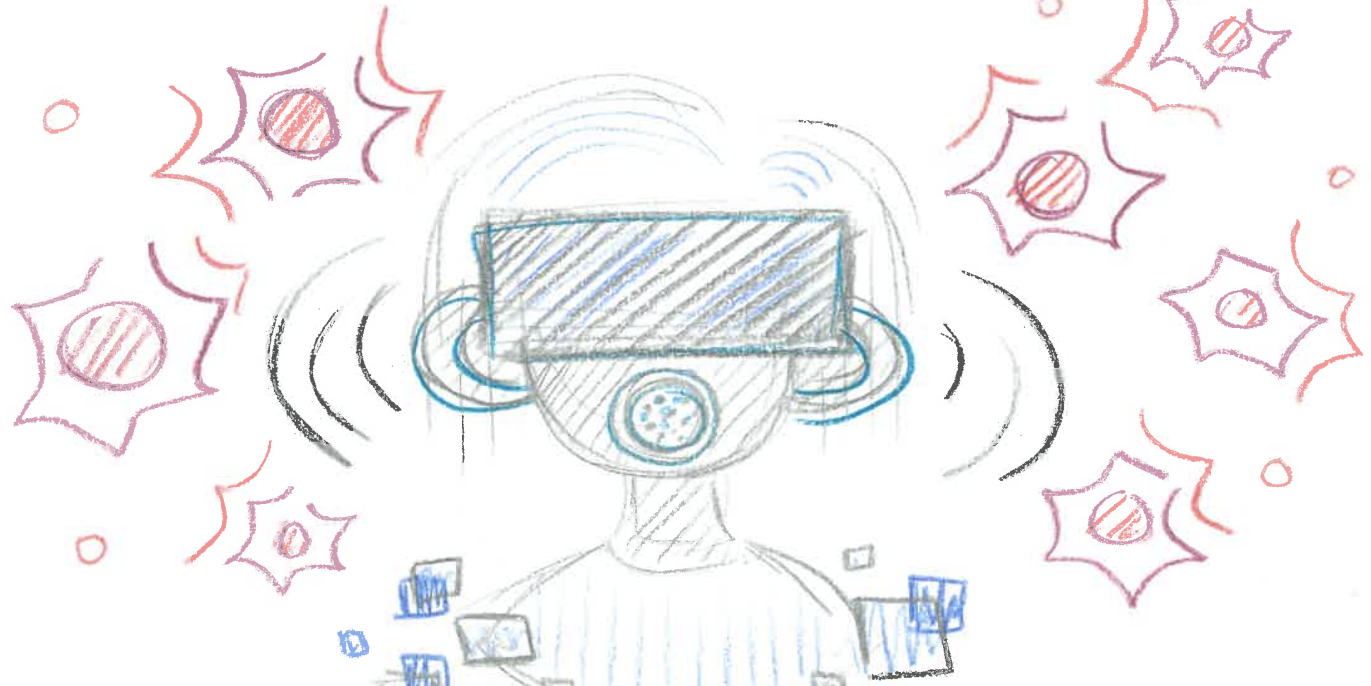}
  \caption{Participant 16 (P16) symbolizes the harms she has experienced as bursting red sparks, which emerge from her personified voice assistant.}
  \Description{}
  \label{fig:P16}
\end{teaserfigure}

\renewcommand{\shortauthors}{Wenzel \& Kaufman}


\begin{abstract}

Voice assistants’ inability to serve people-of-color and non-native English speakers has largely been documented as a quality-of-service harm. However, little work has investigated what downstream harms propagate from this poor service. How does poor usability materially manifest and affect users’ lives? And what interaction designs might help users recover from these effects? We identify 6 downstream harms that propagate from quality-of-service harms in voice assistants. Through interviews and design activities with 16 multicultural participants, we unveil these 6 harms, outline how multicultural users uniquely personify their voice assistant, and suggest how these harms and personifications may affect their interactions. Lastly, we employ techniques from psychology on communication repair to contribute suggestions for harm-reducing repair that may be implemented in voice technologies. Our communication repair strategies include: identity affirmations (intermittent frequency), cultural sensitivity, and blame redirection. This work shows potential for a harm-repair framework to positively influence voice interactions.

\end{abstract}



\maketitle

\section{Introduction}

It is expected that over 48\% of adults in the United States will use voice assistants within the next two years --- that's nearly half of the adult U.S. population \cite{intel2023}. Despite their rapid consumer growth, however, voice assistants still fail to meet the needs of several user groups. Scholars have provided increasing documentation of the ways in which voice assistants underserve diverse populations, including but not limited to women \cite{strengers2021smart}, multilingual and multicultural individuals \cite{choi2023toward}, and people of color \cite{wenzel2023can, koenecke2020racial}. This body of work has revealed how inequitable rates of speech recognition errors lead not only to differences in utility and functionality, but also emotional wellbeing \cite{wenzel2023can}. Users have reported having to code-switch (i.e. alter their speech to a different accent or language variety when speaking with their assistant) simply to have it understand their commands \cite{harrington2022s}. They have also reported instances of disrespect, in which their assistant does not understand the names of important cultural figures \cite{benjamin2020race}. Ultimately, if you are a user who is not prioritized in the design process, it's likely that you may feel frustrated, anxious, and/or angry about your experiences \cite{mengesha2021don}.

In our work, we aim to move past pure documentation of inequities, and begin to consider how these inequities may be readily addressed. To do this, we first capture users' conceptualizations of their voice assistants, including users of Amazon Alexa, Apple Siri, Google Assistant, and Bixby. (\textbf{RQ1:} How do multicultural voice assistant users' frustrations impact their conceptualizations of their voice assistant?) We then create a taxonomy of harms that users have experienced with their voice assistant. (\textbf{RQ2:} What specific harms arise from multicultural users' error-prone interactions with their voice assistant?) Noteably, despite the specificity of this inquiry, the harms we documented have the potential to extend to any user who has the experience of being misrecognized by their voice assistant. Combining all these insights with a strong body of work from social psychology, we create communication repair strategies that address potential user harms. (\textbf{RQ3:} What conversational UX designs may minimize the harm inflicted by error-prone voice assistants?)

To our knowledge, this is the first paper that (1) applies a sociotechnical taxonomy of harms specifically to the socially inequitable aspect of voice assistants and (2) provides communication repair strategies with the unique needs of multicultural users in mind. These communication repair strategies subscribe to an important HCI research agenda toward harm reduction and repair. Through a study with 16 interviews / design workshops with multicultural voice assistant users, our paper makes the following novel research contributions: 
\begin{itemize}
  \item{We provide new insights about how multilingual and multicultural users who are underserved by voice assistants conceptualize and anthropomorphize voice assistants.}
 \item{We outline 6 distinct harms that inequitable voice assistant interactions can inflict.}
 \item{We offer preliminary designs and guidelines for harm-reducing conversational repair that align with both users' preferences and prior research in social psychology.}

\end{itemize}


Our work aims to privilege the voices of users who have historically been excluded from the product development lifecycle. We did this by intentionally focusing on harms that are disproportionately experienced by these users (i.e. quality-of-service harms) and by digging deeper to identify the \textit{downstream harms} of these top-level harm categories. We furthermore selected alternative illustrative methodologies to encourage a more just power distribution between the research facilitator and each participant \cite{packard2008m}, and to help the research community receiving this paper ``adopt someone else's gaze'' \cite{weber2008visual}. 

All users experience errors, but not everyone shares the same subjective error experience. Importantly, we go beyond simply documenting perceptions and taxonomizing harms, and move toward a generative practice to help improve the error recovery experience. The design directions we suggest, while grounded in an investigation with multicultural users, may be deployed for the benefit of all voice assistant users.

\section{Related Work}

We first introduce prior work on how users anthropomorphize voice assistants, as users' relationships with their voice assistants may shape how they interact with and perceive harms. We then review the existing documentation of voice assistant harms, including how voice assistants can yield harmful inaccuracies for certain uses and users. Finally, we introduce the motivation for our harm-repair design, highlighting our unique psychology-based approach. This final section includes a brief overview of psychological harm reduction techniques and reviews how these harm reduction techniques have yet to be realized in existing research.

\subsection{Anthropomorphism of Voice Assistants}
\label{sec:anthro}
Anthropomorphism refers to the tendency to imbue non-human agents with human-like traits, motivations, emotions, and other psychological attributes \cite{epley2007seeing}. A common example in the context of voice assistants is the mindless polite behavior users enact with their assistant (e.g. saying ``thank you'' after their assistant completes a command) \cite{pradhan2019phantom, lopatovska2018personification}. While anthropomorphism can promote user adoption and acceptance of voice assistants \cite{moussawi2021perceptions, wagner2019human}, and has also been tied to user satisfaction \cite{purington2017alexa, shao2021hello}, prior work understanding how users anthropomorphize their voice assistants has been mixed. Research has found some users attribute \textit{positive} human characteristics to their assistant, some attribute \textit{negative} human characteristics to their assistant, and some \textit{do not} anthropomorphize their assistant at all.

Those who hold a positive impression of their assistant perceive their assistant as ``friendly'' and ``helpful'' \cite{park2021exploring, kuzminykh2020genie, schweitzer2019servant}. For users who personify their assistants to a greater extent, they may refer to their voice assistant as a friend or family member, and sometimes even as a girlfriend or wife \cite{gao2018alexa}. More neutral on the spectrum, other users personify their voice assistant as being a ``distant roommate'' \cite{cho2019once}, a librarian \cite{desai2022alexa}, or a professional assistant \cite{kuzminykh2020genie, gao2018alexa}, similar to a master-to-assistant or exchange-based relationship \cite{tschopp2023servant}. In contrast to these relatively positive or neutral social roles, researchers also report on voice assistant users who perceive their assistants as emotionally distant, impersonal, and disingenuous \cite{kuzminykh2020genie, doyle2019mapping}.

Other researchers have challenged the idea of anthropomorphism altogether, showing that users do not personify these agents. For example, in Purington et al.'s analysis of Amazon Echo reviews suggests that while some reviewers do personify their assistant, more than half refer to it as an object \cite{purington2017alexa}, and another small-scale study documented some users referring to their assistant as ``computer'' or ``robot'' \cite{desai2022alexa}. There has also been documentation of users' perceptions of their assistant changing over time. In Cho et al.'s longitudinal study, they found that many users would initially anthropomorphize their assistants, but this tendency would wear off as their assistant's performance fell short of their expectations \cite{cho2019once}. In addition to voice assistant performance, differences in perception are likely based on individual differences in users and study participants themselves. For example, loneliness and social disconnection can increase users' propensity to anthropomorphize their voice assistants \cite{liu2023hanging}. Older adults in particular are more likely to anthropomorphize their assistant and view it as a companion, likely due to the aforementioned cause of loneliness \cite{corbett2021voice, oh2020differences}. In addition, users who live in multi-person households (vs. single-person) are more likely to anthropomorphize Amazon Echo \cite{purington2017alexa, lopatovska2018personification}. This may be due to the fact that smart home assistants reside in an intimate household space where they mediate conversation between household members \cite{voit2020s, beneteau2019communication, porcheron2018voice}.

Of the work understanding anthropomorphism habits, some has focused on specific populations, such as older adults \cite{kim2021exploring} and children \cite{girouard2021children}. Even with this prior work, however, a question still remains regarding how people of color and those who hold a non-American identity may conceptualize their American assistant (e.g., what social roles or attributes, if any, they assign to assistants, particularly given the well-documented lower rates of accuracy and lack of cultural awareness these technologies exhibit).

\begin{itemize}[font=\bfseries,
  align=left,topsep=3pt]
    \item[RQ1:] How do multicultural voice assistant users who have had negative experiences conceptualize their voice assistant? 
\end{itemize}

\noindent  This question is particularly important as these users typically experience lower usability with their devices.

\subsection{Harms of Voice Assistants} 

\subsubsection{High Error Rate, Low Usability}


There are several user groups for whom voice assistants work especially poorly. High error rates have commonly been documented for users who do not speak white American English \cite{ike2022inequity}. Non-native English speakers tend to rate voice assistants lower on usability metrics compared to native English speakers \cite{pyae2018investigating}, and tend to be less satisfied with their experience \cite{pal2019user}. In another study, researchers found that regardless of English fluency, users would still experience greater errors with assistants if they had a ``non-standard'' accent \cite{palanica2019you}. Importantly, voice assistant challenges are not limited to those who speak English as a second language or to those who are multilingual. Native monolingual English speakers can also face a myriad of challenges, and these challenges are especially exacerbated for historically marginalized groups. For example, Koenecke et al. noteably documented a large gap between the error rate that Black Americans and white Americans experience \cite{koenecke2020racial}. 

This disparate error rate may be especially harmful for Black or multicultural individuals who use their assistant for medical information-seeking. It is known that these devices are not optimized for these user groups \cite{harrington2022s}, and voice assistant medication name-recognition has been found to deteriorate for users with non-Anglo American accents \cite{palanica2019you}. These voice assistants have also been documented giving inaccurate information, leading users to false conclusions and inadvisable medical behavior \cite{bickmore2018patient}.

Beyond the explicit bodily harms of a poor medical device, high error rates can lead to two other specific types of harms: psychological harms and quality-of-service harms. Mengesha et al. makes a case for emotional or psychological harm, documenting the lived experiences and feelings of Black voice assistant users through a diary study. Wenzel et al. verified these findings quantitatively, demonstrating in a controlled experiment how voice assistants can negatively affect marginalized users' sense of self-consciousness, self-esteem, and emotional affect \cite{wenzel2023can}. Furthermore, Shelby et al. has identified several quality-of-service sub-harms related to a high error rate, namely alienation (i.e. self-estrangement), increased labor (i.e. the need to exert extra effort or spend extra time), and service or benefit loss (i.e. diminished value) \cite{shelby2022sociotechnical}.

\subsubsection{Existing Harm Taxonomies}
 Shelby et al.'s work highlighted above is one one of the few formal taxonomies that uses voice assistants as a case study. However, voice assistants were not the focus of Shelby et al.'s work, and thus their categorization of harms is not comprehensive. Similarly, Dyal-Chand's powerful \textit{Autocorrecting for Whiteness} outlines several harms that arise from autocorrect, and asserts that these harms may be translated to various other AI technologies \cite{dyal2021autocorrecting}, however researchers have yet to adapt her harm taxonomy. Weidinger et al. proposed a taxonomy of risks for language models, including harms such as discrimination and misinformation \cite{weidinger2022taxonomy}. While this work is no doubt valuable and relevant to the language models that power voice assistants, it does not analyze specific use-cases unique to the voice assistant user experience. Other comprehensive harm taxonomies have been developed for alternative technologies, such as online content \cite{scheuerman2021framework} and online behavioral advertising \cite{wu2023slow}, yet a proper taxonomy and comprehensive harm documentation for voice-based technologies has yet to be developed. 

 Furthermore, beyond what Shelby et al. has contributed, there is little work that identifies the granularities and sub-harms of quality-of-service harms. As users at the margins are typically the ones subject to diminished quality-of-service, it's imperative that we understand precisely what downstream effects they face as a result.

\begin{itemize}[font=\bfseries,
  align=left,topsep=3pt]
    \item[RQ2:]  What specific harms arise from multicultural users' error-prone interactions with their voice assistant?
\end{itemize}

\subsection{Designing for Harm Reduction \& Repair}

\subsubsection{Motivation}
 While the growing body of work dedicated to documenting harms is imperative, there still remains an unaddressed need to develop designs that aid in harm reduction and repair. A common response to the existing disparity in error rate is to create more diverse datasets to train voice assistant language models on \cite{koenecke2020racial, weidinger2022taxonomy}. This solution, however, has a slew of its own challenges \cite{wenzel2023challenges}. Acknowledging the limits and difficulties that arise with improving language models and collecting representative voice data, it is worth exploring ways to reduce harm that go beyond technical improvements \cite{green2020algorithmic}. Our study proposes looking toward social psychology to inform harm reduction in voice assistant communication repair.

\subsubsection{Lessons from Psychology}
We begin by building upon recent work that positions voice interaction failures as microaggressions \cite{wenzel2023can}. Prior research in psychology has revealed that experiences with microaggressions can trigger stereotype threat, the anxiety that emerges from the fear of being the target of a stereotype or of unintentionally conforming to a stereotype \cite{steele1995stereotype}. The experience of stereotype threat can be associated with a number of debilitating cognitive and affective outcomes, including: heightened self-scrutiny and self-reproach; reduced self-esteem; and threats to a sense of safety and belonging \cite{shapiro2007stereotype, beilock2007stereotype, woodcock2012consequences}. At the same time, the impact of threat can be reduced by a number of methods, such as actively negating or dismissing an activated stereotype \cite{nadler2011stereotype, kawakami2000just}; increasing the accessibility of a positive aspect of one’s identity to counteract a negative stereotype \cite{rydell2009multiple}; or affirming one's values as means of reinforcing one's self-integrity and providing a buffer to the psyche against threat \cite{martens2006combating}. Research on interventions to mitigate the effects of microaggressions and stereotype threat have found that they are more effective if implemented in subtle ways (i.e., in ways that do not position an individual as vulnerable or weak) and, at the same time, are asset-based (i.e., leverage personal strengths as a means of empowerment) \cite{samuelson2016community}. However, work studying how these interactions may be mediated through conversational agents, let alone technologies in general, is limited.

\subsubsection{Communication Repair Interfaces \& Interactions}
 Before conducting a redesign that integrates psychological theory, we first require an understanding of the current state of communication repair. Cuadra et al. found that if a voice assistant recognizes and repairs its mistake, users will evaluate it more favorably \cite{cuadra2021my}. In a similar vein, Mahmoud et al. concluded that voice assistants that assume blame and apologize sincerely for their mistakes are perceived to be more intelligent, more likeable, and more effective in their error recovery \cite{mahmood2022owning}. Kim et al. explored different methods for communication repair for vehicular voice interfaces with mixed results. Overall, they recommend using brief natural commands that include an explanation or status report to help minimize confusion for users \cite{kim2019did}. 

 Notably, the work summarized here does not focus on users at the margins of the voice assistant user experience. As an increasing body of literature has come to demonstrate the unique experiences and needs of marginalized voice assistant users and users from diverse cultures \cite{wenzel2023can, mengesha2021don, choi2023toward}, it has also become increasingly important to understand how communication repair should be adapted for these specific populations, as they are more likely to experience errors and conversational breakdown.




\subsubsection{Existing Design Research for Diverse Users}
 Researchers have also highlighted contrasting design needs between non-native and native English speakers. For example, non-native speakers appreciate longer turn-taking time, whereas native speakers find long turn-taking time disruptive \cite{wu2020see}. Other work on multilingual users suggests that future voice assistants should be designed to \textit{code-mix}, i.e. be capable of alternating between different languages within an interaction \cite{choi2023toward}. 

\begin{itemize}[font=\bfseries,
  align=left,topsep=3pt]
    \item[RQ3:]  What conversational UX designs may minimize the harm inflicted by error-prone voice assistants?
\end{itemize}

\section{Methodology} 
\begin{table*}[h!] 
\caption{ Self-reported demographics of all participants. Languages are listed in order of dominance (based on LEAP-Q). \textit{Note:} Many participants had experience with multiple voice assistants. For the purpose of the study, we asked participants to focus on their experiences with a single assistant. This single assistant is what is noted in the fourth column. (P6 is an exception.)}
\begin{tabular}{|c|l|l|l|}
\hline
\textbf{ID} & \textbf{Languages}                        & \textbf{Self-Identified Cultures}                                 & \textbf{Assistant}               \\ \hline
P1          & English                                   & Indian, US-American                               & {Amazon Alexa}                                                       \\ \hline
P2          & English, Hindi, Tulu, Kannada             & Indian                                            & {Google Assistant}                                                   \\ \hline
P3          & English, Mandarin                         & US-American, Chinese, Queer                       & {Apple Siri}                                                        \\ \hline
P4          & English, Spanish, Indonesian              & US-American, Chinese-Indonesian, Ashkenazi Jewish & {Google Assistant}                                                        \\ \hline
P5          & English, Korean, Spanish, Japanese        & US-American, Asian American, Korean               &  {Samsung Bixby}                             \\ \hline
P6          & Hindi, English                            & Indian                                            &  {Alexa \& Google}                                                      \\ \hline
P7          & English, Korean, Spanish                  & US-American, Black-American                       &  {Google Assistant}                                                        \\ \hline
P8          & Hindi, English                            & Indian                                            &  {Google Assistant}                                                        \\ \hline
P9          & English, Hindi, Sindhi                    & US-American, Indian, Sindhi                       &  {Apple Siri}                                                        \\ \hline
P10          & Marathi, English, Hindi, German, Japanese & Indian                                           &  {Google Assistant}                                    \\ \hline
P11         & English, Tamil, Hindi                     & Indian                                            &  {Apple Siri}                                    \\ \hline
P12         & English, Mandarin                         & US-American, Chinese                              &  {Apple Siri}                                   \\ \hline
P13         & Malayalam, Hindi, Tamil, English          & Indian                                            &  {Apple Siri}                                    \\ \hline
P14         & English, Spanish                          & US-American, Mexican                              &  {Apple Siri}                                   \\ \hline
P15         & English, Korean, Mandarin                 & Korean, Taiwanese, US-American                    &  {Apple Siri}                                   \\ \hline
P16         & English, Urdu/Hindi, French               & US-American, South Asian, Muslim                  &  {Amazon Alexa}                                  \\ \hline
\end{tabular}
\label{table:demographics}
\end{table*}

 \subsection{Recruitment} All materials and procedures described below were approved by the institutional review board at the authors' university. A total of 16 participants were recruited through flyers posted around Pittsburgh, Pennsylvania. Flyers were pinned at 2 local university campuses, at public libraries, public transportation stations, and at local cafes and shops. Participants were screened for eligibility, and were required to have had prior negative experiences with voice assistants as assessed through a scale adapted from the Language Experience and Proficiency Questionnaire (LEAP-Q) \cite{marian2007language, kaushanskaya2020language} (``When speaking to your voice assistant in [self-identified language], how frequently does it misunderstand your requests?''). Participants were also required to be aged 18 years or older and were asked to self-identify their culture(s) and language(s) they spoke, also using questions from LEAP-Q, (See Table \ref{table:demographics}). We prioritized inviting participants who identified with more than one language or culture. All participants were compensated with an Amazon gift card worth USD 20.

 \subsection{Procedure} 
Prior to formal data collection, the study procedure was piloted with 10 participants from the target population in order to get an accurate understanding of the participant time commitment, ensure that the study procedure was clear and had a coherent flow, and to ensure the study activities captured appropriate data. 
 
\subsubsection{Data Collection} \label{datamethod} The study was conducted in a lab at the authors' university campus. Each study session lasted for 40-60 minutes, depending on how long the participant took to complete the study tasks. Upon entering the room, the participant was asked to read and complete a paper consent form. The participant then engaged in a directed storytelling activity in which they were asked to describe a frustrating experience they had with their voice assistant. Participants who had experiences with multiple assistants were asked to focus on a single assistant. Methodologically, having each participant focus on a single assistant allowed for internal consistency across study activities. Practically, focusing on a single assistant ensured that the research activity would take no more than an hour of a participant's time. (An exception was made for P6, who weaved his experiences with both Amazon Alexa and Google Assistant into a personal narrative.) Based on the participant's story, the first author (PI) led a semi-structured interview. The directed story-telling and semi-structured interview aimed to partially address RQ1 in addition to largely focusing on RQ2. 

The second activity asked the participant to draw a personified version of their voice assistant, and then describe the drawing to the PI, a method similar to that used in prior work \cite{kuzminykh2020genie, curry2020conversational}. In addition to helping us effectively address RQ2, this participatory visual method helped the researchers renounce some of their power and cultivate a collaborative research ethos through mutual discovery \cite{packard2008m}.

The third activity asked the participant to write a love or breakup letter to their voice assistant \cite{hanington2012universal, gerber2011tech}. This letter-writing activity revealed both participants' conceptualizations of their assistant and harms that they had experienced, addressing both RQ1 and RQ2. 

The fourth and final activity was generative in nature, and intended to address RQ3. It asked the participant to write messages that they wished their voice assistant would say when an error occurred (i.e. conversational repair). The participant was asked to write four affirming messages (i.e. messages of positivity and warmth, or cultural messages) and two ``freeform'' messages (i.e. messages without any formatting or content requirements). After each of the interactive activities, participants were invited to elaborate and explain their illustrated/written choices.

P1 through P5 had the PI present for facilitation and two research assistants present for notetaking. P6 through P16 only had the PI present for the study, and had their sessions audio-recorded with consent. There was no observed difference in participants' self-disclosure across the two facilitation styles. 
 
\subsubsection{Data Analysis} Data analysis was conducted through both inductive and deductive coding by the PI. For the inductive open coding, initial descriptive codes \cite{wolcott1994transforming} were applied which helped note relations between the topics which emerged from participant quotes and the three research questions. These typically corresponded to the intended activity as noted in Section \ref{datamethod}. Topics mentioned by participants that did not directly relate to a research question received their own unique codes. Sub-codes \cite{saldana2021coding} were then applied to denote topical results to the research questions. (See Table \ref{tab:subcode} for coding example.) The visual and written data we collected were coded alongside participants' verbal descriptions. Their verbal descriptions heavily aided the interpretation of the data, so as to avoid overinterpretation of the visual materials. Overall, there were 13 miscellaneous codes (not directly related to an RQ), 6 sub-codes that related to RQ1, 6 sub-codes that related to RQ2, and 15 sub-codes that related to RQ3 Additional deductive rounds of coding using the taxonomy of harms developed by Shelby et al. \cite{shelby2022sociotechnical} and Dyal-Chand \cite{dyal2021autocorrecting}, respectively, were also completed to address RQ2. All codes received a frequency count. 

\begin{table*}[h!]
    \centering
    \begin{tabular}{|l|l|l|}
    \hline
        \textbf{{Participant Quote}} & \textbf{ {Descriptive Code}} & \textbf{ {Subcode}} \\ \hline
         {``It kind of feels like...somebody who has no social skills''} &  {VA} &  {unsociable} \\ \hline
    \end{tabular}
    \caption{ {Example code, including quote which received the code \textit{``VA - unsociable.''} VA = Voice Assistant and denotes the relation to RQ1, which inquires about participants' voice assistant conceptualizations. Unsociable describes how this participant conceptualized their assistant. (See Section \ref{sec:unsociable} for the results related to this code.)}}
    \label{tab:subcode}
\end{table*}
 

 {\section{Findings}}

 This section is broken down to first describe participants' perceptions of their voice assistant, as analyzed through their drawing and letter-writing activity. It then goes on to outline participants' stories and pain points. Lastly we move on to describe participants desired conversational repair messages.

\subsection{A Social Agent with ``No Social Skills'': How Users Conceptualize Their Voice Assistant}

\subsubsection{Knowledgeable but Incompetent} 
\begin{figure}[h]
\centering
\includegraphics[scale=.3]{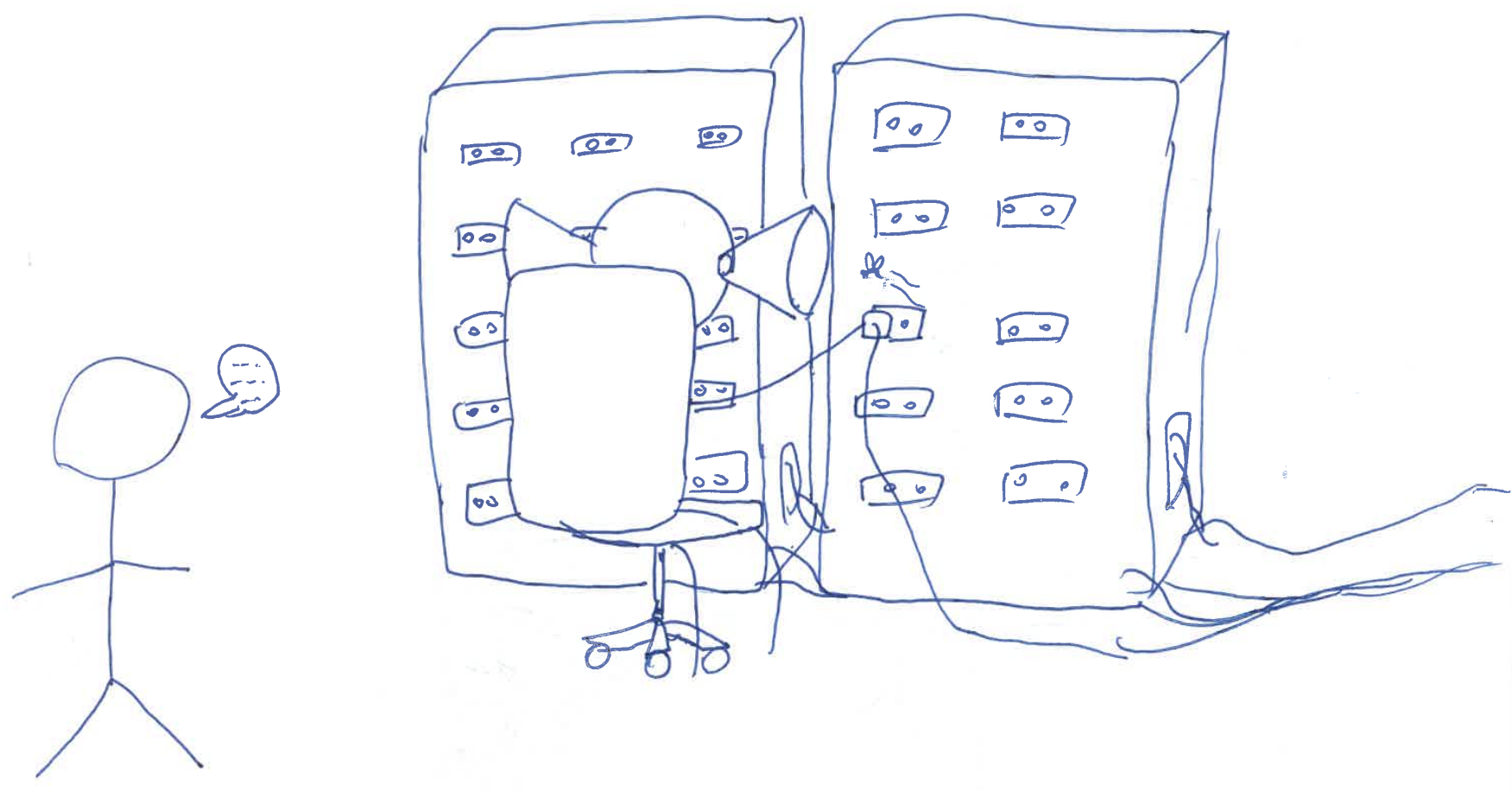}
\caption{\textbf{Participants often envisioned their assistant as being low-tech.} P15 was one of the few participants who incorporated technology in their conceptualization of their assistant, and he did so by including a large, outdated \textit{``analog call center.''} The lefthand stick figure represents P15, and the righthand figure represents Siri. To process requests, Siri must use the button-operated machine. Siri is faced away from the P15 \textit{``because it doesn't really take into consideration what I was doing.''} A fruit fly circles around Siri, demonstrating how Siri may be distracted and perform poorly.}
\label{fig:P15} 
\end{figure}

Many participants described their assistant as  {intelligent yet incompetant}, highlighting the limits of their assistants' abilities. This largely manifested as the participants describing their assistant as incredibly knowledgeable (``smart''), but lacking in their communication and task execution abilities (``dumb''). This insight emerged in a few key ways. Firstly, many participants either drew their assistant with glasses, as an archetypal indication of intelligence, and/or as a secretary \textit{``because secretaries tend to know everything about the area''} (P14). Simultaneously, however, their assistants were also depicted as being low-tech. In their sketches, they noted their assistants working with many paper documents and pens and lacking technological equipment. Those who did depict their assistant with technology made sure to express their assistants' limitations in other ways. For example, P15 included oversized manually-operated machines in their drawing (Figure \ref{fig:P15}) as opposed to modern devices. P15 went on to describe that the personification of his ideal assistant would be a person working with a laptop, implying that the current assistant is not advanced enough for his needs. In contrast to P15, P6 depicted his assistant using modern technology (a tablet), but his overall sentiment was still the same as P15. He described how in spite of having all possible resources, his assistant still failed to actually assist him: \textit{``it has all the tools, it has everything, probably everything in their arsenal, but [it's] not able to help me.''} In a related mental model, P11 expressed these limitations by drawing his assistant Siri over a hill in the distance. He explained that because of its technical limitations it feels like Siri is \textit{``far from me, and not next to me.''} Ultimately, regardless of how they chose to depict their assistant, it was clear that participants believed their assistants to be knowledgeable, but unable to properly harness their knowledge.


\subsubsection{Lacking Emotional and Social Skills} \label{sec:unsociable}
\begin{figure}
\centering
\begin{subfigure}{.4\textwidth}
  \centering
  \includegraphics[width=.5\linewidth]{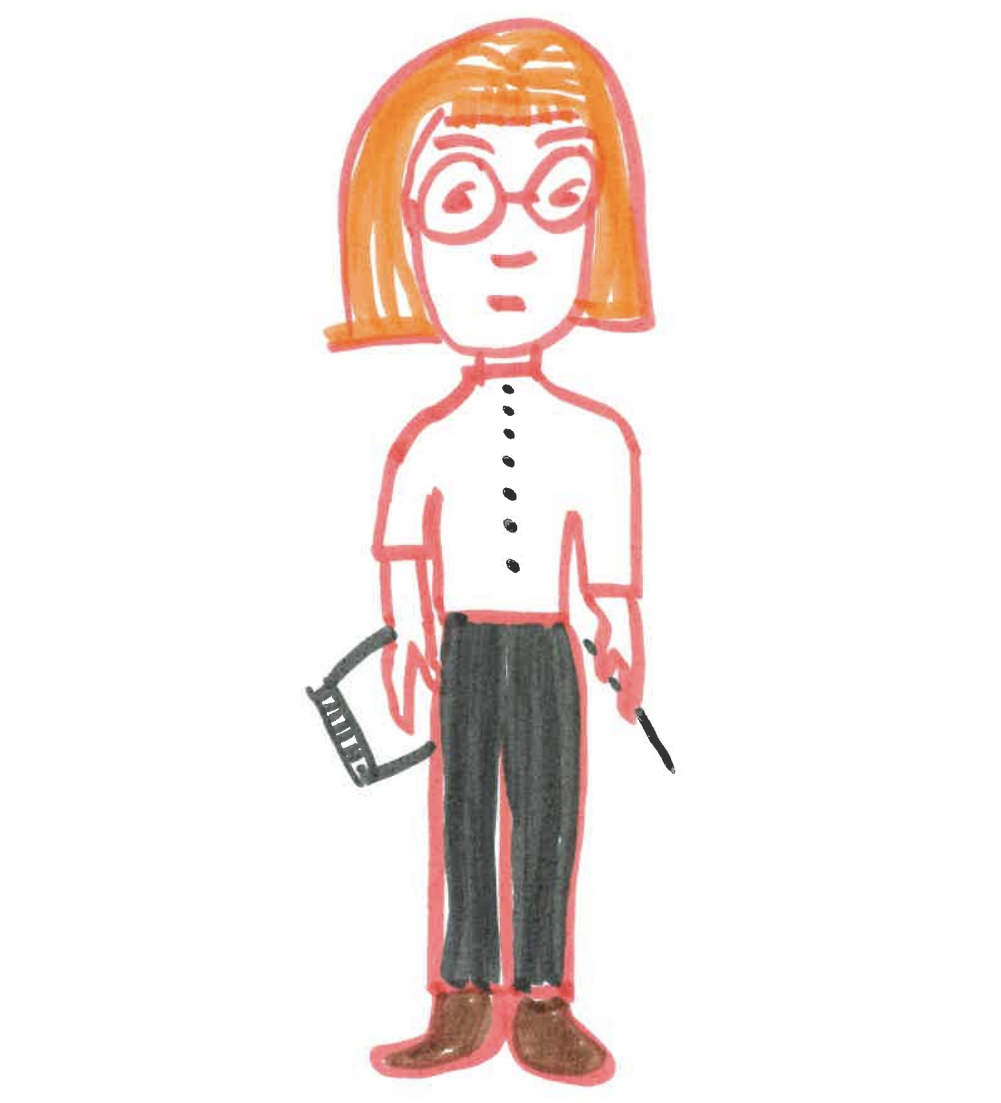}
  \caption{For P12, Siri has a (low tech) notepad and pen because she \textit{``takes notes and remembers things.''} Regarding her emotions and personality, Siri has \textit{``no feeling or expression''} and has \textit{``nothing frilly or artistic or exciting.''} }
  \label{fig:P7}
\end{subfigure}%
\hfill
\begin{subfigure}{.5\textwidth}
  \centering
  \includegraphics[width=.7\linewidth]{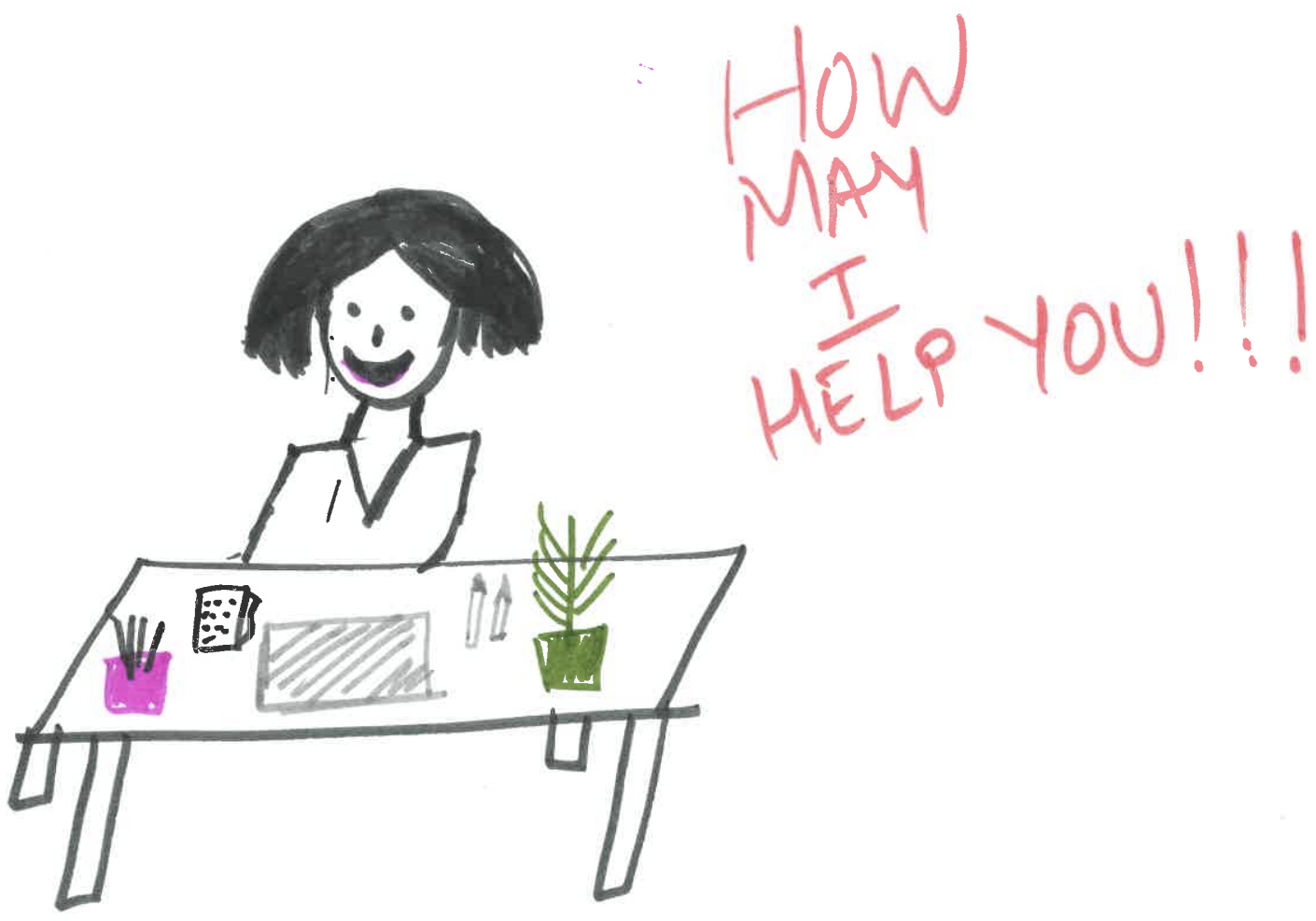}
  \caption{This drawing captures P6's embodiment of both Alexa and Google Assistant \textit{``because ultimately, at the end of the day, I'm getting frustrated by both of them.''} P6's assistants have \textit{``all the tools''}, as conveyed in through their desk setup, yet still \textit{``no matter how many times she may ask, HOW MAY I HELP YOU?, she won't understand what I'm trying to say.''}}
  \label{fig:P6}
\end{subfigure}%
\label{fig:emotion}
\caption{\textbf{Participants perceived their assistant as lacking emotional and social skills.} This manifested either as (a) lack of emotion or (b) possessing fake emotions and overstepping communication norms.}
\end{figure}

 This perceived incompetence often translated to a lack of emotion. In many participant drawings, the voice assistant appeared unfriendly and emotionally cold, with a straight face. P7 described how Google Assistant is \textit{``always very serious''} and \textit{``would not be a very light hearted person''} (Figure \ref{fig:P7}). In a similar vein, P10 describes their assistant as someone who is \textit{``grumpy''} and displays little emotion. P10 also stressed that her assistant is \textit{``somebody who has no social skills and doesn’t understand the details of social situations.''} She emphasized how she associated the lack of facial expression in her drawing with her assistant's \textit{``dumbness.''} Multiple participants made a connection between their assistant and real people they knew in their life, citing coworkers who \textit{``cannot be creative''} (P12), friends who are oblivious and \textit{``believe [sarcasm] is true''} (P10), bad listeners who require you to \textit{``shout at them...two to three times''} (P3), and acquaintances who are an introverted  \textit{``homebody''} who \textit{``know random facts''} (P14). When participants did depict their assistant as emotionally expressive, it was often with an artificial fervor. For example, P6 depicted his assistant with an overzealous smile, annotated with a monologue of ``HOW MAY I HELP YOU!!!'' (Figure \ref{fig:P6}) While these two emotional depictions are apparent opposites (lack of emotion vs. overexpression of emotion), they are similar in that they are both extremes representing atypical social behavior.

\subsubsection{Contending with the (non)human} \label{sec:nonhuman}
 Participants' difficulties conceptualizing their assistant often stemmed from a confusion about their assistants' human-like nature. As many scholars have demonstrated in the past, people unconsciously attribute human characteristics to non-human entities, and accordingly treat non-sentient entities as humans \cite{nass1994computers}. This phenomenon extends to voice assistants \cite{schneider2022assisting, mclean2019hey, wenzel2023can}, and several participants expressed awareness about their miscalculated expectations. P16 describes: \textit{``When I correct people in real life, they learn! They fix whatever information they had before and don't repeat the mistake!''} P12 explains in more detail, saying: \textit{``I’m just used to talking to a person and having a person understand. And perform a function just smoothly, no issues. And I can’t understand why a computer that’s like been made to work in a similar fashion just cannot with the easiest functions. So I think getting used to working with people who are very good listeners, and understanding, and just know how to do the task spoils me.''}.



\subsubsection{Device Matters}

 Several participants had experiences with multiple voice technologies, and a consistent trend across all of these participants is that they were overall less satisfied with Siri than with other assistants. These participants explained that they had switched phone providers (i.e. Android to iPhone), for reasons unrelated to the assistant experience and as an unintended consequence their voice assistant experience suffered. P9 begins his letter to Siri, \textit{``Dear Siri I’m done done. I’m only tied to you because of Apple. But would switch to Google Assistant the day Apple allows me to switch. You talk nonsense half the time, and are unable to do what I want.''} He went into greater depth through his drawings, explaining that Google Assistant \textit{``is way more sophisticated and well presented...in terms of dressing this would look like an investment banker. And listens to Beethoven-style like professional music.''} In contrast, his idea of Siri is \textit{``dressed as a Silicon Valley hippie youngster, who listens to rock and roll and is way more simple [minded].''} These archetypes are reflective of the perceived competence and performance accuracy of the two assistants. While this difference in evaluation across assistants is not our primary finding, it has important implications for how researchers should approach future studies on voice assistants. 

\subsection{Harms of Encountering Biases in Voice Recognition Systems}

\begin{table*}[t]
\begin{tabular}{|p{3cm}|p{7cm}|p{3.7cm}|}
\hline
\textbf{Harm}         & \textbf{Example/Definition}                                                                                         & \textbf{ {Related Documentation}}                                                   \\ \hline
Relational Harm           & Interpersonal conflict (that likely would have been avoided with a properly functioning device)                     & Storer et al. \& Voit et al.                                                               \\ \hline
Service/Benefit Loss  & ``paying the same price for a less useful product'' \cite{dyal2021autocorrecting}                                        & Shelby et al. (\& Dyal-Chand as ``Economic Harm'')                    \\ \hline
Increased Labor       & Exerting extra effort (i.e. by being forced to revert to manual methods, modifying one's speech, repeating oneself) & Shelby et al.                                                     \\ \hline
Identity \& Cultural Harm   & Invalidating users' cultural identity (i.e. by not recognizing a important names or proper nouns)                   & Dyal-Chand (encapsulates Shelby et al.'s ``Alienation Harm'')         \\ \hline

Physical Endangerment & Users are put in physical danger without the presence of a malicious actor, (i.e. user at risk of crashing while driving)                              & Bickmore et al.                                                               \\ \hline
Emotional Harm            & Emotional distress (i.e. lowered mood, worsened self-esteem)                                                        & Shelby et al. (nested under ``Diminished health and well-being'') \\ \hline
\end{tabular}
\caption{Table of 6 voice assistant harms related to quality-of-service harms. Related Documentation references studies with similar harm documentation. While similar, however, the documentation is not completely identical. Please see Section 6.3.}
\end{table*}

\subsubsection{Relational Harms}

 Multiple participants described instances of misattributed blame, in which they were blamed for a problem that was actually the fault of their voice assistant. This experience often led to interpersonal conflict. For example, P8 expressed frustration regarding how Google Assistant incorrectly inputted meetings in his calendar. \textit{``I missed some important meetings, and it was I who was blamed for missing those meetings.''}\footnote{In this context, considering it was P8's professional working relationships that were harmed, taken further, this issue of misheard voice commands could grow into an economic harm or an allocative harm.}

 Of course, non-professional relationships are also at risk. P13 explained how his grandmother began using Siri to make phone calls, as the iPhone interface was difficult for her to navigate. This led to a barrage of calls from his grandmother to unintended recipients, in addition to random calls throughout the night. P13 fervently explained his family's deep frustration with his grandmother: \textit{``We wouldn’t directly question the technology because we would say `okay, this has to work. If you’re doing it correctly, this has to work.' So we started pressurizing her like `okay, you are not saying it right. You have to say it right!'''} Later, in a visit to his grandmother, they learned that she was, in fact, using the correct voice commands, but it was Siri who was directing calls to the wrong people.

\subsubsection{Service/Benefit Loss}
 In both of the stories outlined above, the conflicts that arose from these conversational breakdowns led participants to revert to manual methods. P8 started using a paper journal to keep track of his meetings, which he felt was more reliable than voice-based or digital methods. P13's family replaced his grandmother's touchscreen iPhone with a flip phone and taught her to use speed dial. This regression towards older, more manual, methods echoes economic harms outlined by Dyal-Chand: ``While we may also suffer from losing a job...'' (reflective of P8's struggle), ``the core economic harm here is from paying the same price for a
less useful product.''\footnote{Note that \textit{Economic Harm} as defined by Dyal-Chand is different from the \textit{Economic Loss} harm defined by Shelby et al. It instead more accurately reflectes \textit{Service/benefit Loss} as defined by Shelby et al.}

 Interestingly, despite the frustrations all participants faced, when describing their ideal assistant, they did not ask for one that was completely fluent or multilingual. P9 said \textit{``it works perfectly fine,} if \textit{you understand that you’re talking about a machine.''} Similarly, when describing the troubles he has communicating in Spanglish using the voice-to-text feature, P14 was sympathetic to the limits of the technology: \textit{``it's not its fault, I guess, because it's set to English.} However, he still voiced some desire for a multilingual assistant, even if it knows only \textit{``a little bit''} of Spanish. P13 explained that he believed a multilingual assistant seemed too difficult to develop, and instead suggested that assistants should simply advance towards recognizing different accents. He qualified this statement further, saying that assistants \textit{``should be able to pick up accents to some extent, not to 100 percent, but to some extent.''} Users have been underserved by their assistants for so long that they've fully recalibrated their expectations. They \textit{expect} to buy products that do not serve them. They do not even outwardly wish for products that can fully understand them.

\subsubsection{Increased Labor}
 The high disparate accuracy rate leads to ``unequal access to and through technology'' \cite{dyal2021autocorrecting}, which can often lead to what Shelby et al. term as an ``increase labor'' harm. While users who benefit from voice assistants may experience greater efficiency and convenience in their day-to-day workflow, others must exert extra effort simply to attempt to have their assistant understand them. Even after going through this extra labor, the assistant is not guarranteed to function properly. P16 echoes this sentiment well, demanding angrily in her break-up letter to her assistant: \textit{``Why should I have to} repeatedly \textit{bend over backwards just to get you to understand what I'm saying? How come my all-American named friends get treated better than I do? Why is convenience only allowed for some people?''}

\subsubsection{Identity \& Cultural Harms} \label{sec:culture-harm}
 In spite of all our participants being multilingual, only a subset of them have tried using non-English languages with their assistant. More common, however, was the phenomenon of codeswitching, in which participants would alter their style of speech to be closer to a white American English speaker. P16 described her experience of trying to call family members, saying that in order for it to work correctly \textit{``we have to deliberately mispronounce [their names],''} stripping away her family's culture an identity. P10 gave a clear example, alternating between different voices, stating: \textit{``I also feel dissatisfied in one particular way that when I say `OK, Google,' it does not understand my accent. And when I say it in a different accent, like,} `OK, Google,' \textit{that’s when it understands.''} Users must assimilate to a standard white American middle or upper-middle class accent, effectively altering their identity, just to be recognized by their voice assistant. Even then, they are not always successful because \textit{``how good your fake accent is also matters''} (P13).

 While accents are indeed a powerful reflection of one's identity, there are other ways identity and cultural harms can manifest. Vocabulary also matters. P1 described his frustrating experience trying to add Indian spices to his online shopping cart, to no avail. He explained he tried adding it both by their English translation and their Hindi name. When contemplating what he could do to solve his problems with Alexa, he suggested \textit{``Maybe I should try learning words which could act like a replacement...or maybe the word I had in my mind is not the right word for that item.''} --- In spite of already trying two of the most popular languages in the world, he was still convinced his commands were not good enough. His final suggestion was that he should alter his identity with \textit{``accent training.''}

 P7 shared their experience growing up in a Black household, and noticing how their experience with technology changes as their cultural language gets adopted by the young American majority: \textit{``I just use a lot of Black slang, a lot. And usually, [my voice assistant] does not know what those things mean until it becomes a mainstream word on the internet.''}


\subsubsection{Physical Endangerment}
 Unique from Shelby et al.'s taxonomy's ``Technology-facilitated violence'' and Storer et al.'s ``Physical Safety'' tradeoff, we highlight that violence and physical danger are not solely a byproduct of a malicious actor taking advantage of a system. There are indeed cases where users, against anyone's own desires, are put in danger due to a faulty system  that has been promoted without great consideration for a diverse user base. A clear example of this was in the case of driving and navigation. While prior work has highlighted car navigation as preferred use-case of voice assistants, yielding higher ratings of usefulness, trust, and positive emotions when compared to alternative tasks \cite{he2022battle}, our results suggest that this is likely not the case for marginalized users. P12 described \textit{``Oftentimes, I feel like I’m pronouncing things very clearly and loudly, but it still can’t understand me. And I don’t know what’s going on. And I don’t know where I’m going. So it’s just this, this frustrating experience, and very dangerous and confusing.''} Other participants echoed these feelings of fear, danger, and uneasiness when using navigation apps. Importantly, in such scenarios, the drivers using voice-interaction are not the only ones in danger. Rather, there are many stakeholders (i.e. passengers and other drivers on the road) for whom their safety is put at risk.

\subsubsection{Emotional Harms}
 While all harms have the potential to cause some level of emotional distress, there were cases where negative emotions, like sadness or deflated self-esteem, were particularly salient. Such was the case for P14. He explained, in the context of sending text messages with bad news, when his assistant fails to transcribe his voice it \textit{``adds to the frustration just like, `Man, I gotta deal with this,' like little thing after a little thing. It builds up.''} The simple frustration of a conversational breakdown can exacerbate already negative emotions. Recalling prior work on the intersection of race and conversational breakdowns \cite{wenzel2023can}, we'd like to reiterate that such frustrations are likely to be exacerbated for people of marginalized identities, due to the emotional and cognitive impacts of microaggressions.

 Importantly, harms are not mutually exclusive. It is not unlikely that when someone experiences, for example, an Identity \& Cultural Harm that they also experience emotional harm as well. P16 describes how misunderstandings affect her through a range of different emotions: \textit{``It definitely makes me feel a little bit alienated worst case scenario, and mildly annoyed, best case scenario, neither of which are very fun.''} Negative emotions may even make other harms more potent, heightening their effects, and thus should be taken seriously. 

\subsubsection{Persistent Harms Across Non-English Devices}
 While significant prior work has focused on English-based voice assistants, a subset of our participants described experiences of using voice assistants with alternative default languages. Their experiences showed that accurate speech recognition is a problem that persists across various agents. Similarly to how those who speak English as a second language had trouble using English-based assistants, P7, P15, and P5 described their struggles speaking Korean as a second language to Korean assistants. Even those who are native speakers have trouble using voice technologies. P13 and P5 described challenges their families faced speaking with native regional accents to devices programmed for Hindi and Korean, respectively. In these cases, the language ``ideal'' is not that of a white American, but of a person who is of a high socioeconomic class from that country.

\subsection{Users' Desired Conversational Repair Strategies}

In the following section, we analyze the final task of the design workshop, in which participants were asked to write messages they wish their assistant would say in the face of communication breakdown. Participants were instructed to write both \textit{freeform} messages and \textit{affirming} messages (see Methodology \ref{datamethod}).

\subsubsection{Freeform Messages} Without any specific guidelines, participants' freeform messages typically revolved around improving overall usability and convenience, rather than positive harm-repair techniques.

\paragraph{Understand Context}
Participants voiced a desire to have their assistants be context-aware. For example, P8 suggested that the assistant should understand what social context users are in (ex. based on the surrounding decibel level). He described how if the assistant was able to detect whether the user was in a social, work, or solo environment it could adjust its sensitivity to commands, to ensure it does not activate inappropriately given the specific social environment.

\paragraph{Keep it Short} Participants explained that affirmations seem nice, but in practice they may prefer shorter, briefer interactions. This follows suit with voice assistant design affordances, as one major user experience goal is to offer convenience to users. One strategy participants voiced to accomplish this is to have the assistant specify which keyword from the command it has low confidence in (ex. if asking to set a reminder about lunch at noon, and the assistant does not capture the entire command, it might ask ``For what time should I set the lunch reminder?'' instead of asking the user to repeat the entire phrase.) Another strategy several participants suggested is having the assist offer options (ex. ``Should I set the reminder for 10AM? or 10PM?''). By giving users options to choose from, users are not required to repeat an entire phrase and the verbal interaction can be shortened.

\subsubsection{Affirming Messages}
When asked to write affirming messages they would like to see from their voice assistant in response to errors, participants imbued more positivity and compliments into their error messages, i.e. \textit{``Thank you for your patience! This is what I think you mean...''} (P3) or \textit{``That sounds really interesting. Do you think you could say that one more time?''} (P14). Some participants made sure to employ more polite language, i.e. \textit{```I'm very sorry' instead of just `I'm sorry'''} (P15), while others chose a more casual tone or even involved humor, i.e. \textit{``Jeez, I must be having a case of the Mondays, because I think I missed that. What did you say?''} (P4). The latter two examples notably involve the assistant taking responsibility for its mistake, tying into design considerations surrounding blame.

\paragraph{Blame \& Ownership}
In the semi-structured interview, a few participants highlighted feelings of blame and shame with respect to their errors; (see P1's thoughts in \ref{sec:culture-harm}). Perhaps in response to this phenomena, when writing desired error messages, participants focused on incorporating apologies and blame attribution. P13 wrote \textit{``Hey [Name], unfortunately I couldn’t get all of it, so would you mind repeating that? I am sorry.''} Here, the assistant uses the pronoun ``I'' to admit that it is at fault for misunderstanding the user's command. In another example, P2 wrote \textit{``I might have misheard you, but here are the responses closest to what I think you meant.''} In this instance, the assistant qualifies its response by letting the user know that it may have improperly captured the user's request. Noteably, these messages do not tell the user that they spoke incoherently. Rather, these messages place emphasis on the assistant's faulty reception of commands. While these responses may not have a particularly positive valence, like some of the other affirmations users wrote, they help affirm users by allowing the assistant to assume the blame.  

\paragraph{Affirmation through Cultural Awareness} \label{sec:aware}
Participants who invoked culture in their responses often did so in subtle ways, referencing art, sports, and pop culture. For example a response crafted by P7 referenced their artistic identity by complimenting their music taste:  \textit{``Your taste in music is amazing. But I don’t think everyone is ready for that, because I can’t find it.''} They explained how this response was relevant as music-playing and music-searching commands were their top use cases. Another participant, P10 pointed to one specific message she wrote and proclaimed \textit{``if it says only this one and nothing else, that would be good for me [giggles], because it's [a quote] from my favorite YouTube channel.''} These pop cultural references are notably not related to the users' cultural heritage, in spite of that being a key factor in our recruitment. In fact, some participants even made negative remarks about the potential for their assistant to make comments about their cultural heritage. Such comments were perceived as an invasion of privacy. Participants who associated their assistant with its corporation (i.e. Apple, Amazon, Google) also found cultural heritage references to be uncanny due to the idea of corporate inauthenticity (see next subsection). 

There were, however, still cases in which participants referenced their cultural heritage tastefully. For example P9 described how when he code-mixes his commands (i.e. combines two languages within a single command) his assistant always defaults its interpretation to English. Understanding that assistants may not be capable of code-mixing yet, he suggested that his assistant could ask \textit{``Okay, can I switch you to [a regional language]?''} This response is affirming because it acknowledges a specific part of his identity, rather than forcing him into use American English. P2 gives another example of an indirect cultural acknowledgement that could be used as an error response: \textit{``Sorry, I didn’t quite catch that, just like Sachin Tendulkar in last week’s cricket match.''} This message is not overly polite nor is it overly positive. Instead, it mixes humor with cultural knowledge, and in doing so it acknowledges the user's identity. Participants noted that these error responses could utilize regional geolocation data, and that this data felt broad enough to not be privacy-invasive.

Another participant (P5) stressed that affirmations must be culturally sensitive. She explained how in the United States, affirmations are often tied to a person and their individual characteristics, whereas in Korea, they are more likely to be tied to an action or behavior a person exhibits. Giving a basic example, she said \textit{```Wow you’re such a loving person.' --- If someone told me that, I'd be like `What?!' That doesn’t makes sense to me. It would take me out of it.''} P5 contrasts this with action-based examples; for a voice assistant error message, she wrote: \textit{``I'm glad you're exploring healthy food options. What was the recipe you wanted to hear about, again?''} Here, you can see that the praise is directed toward the action of searching for healthy food options, as opposed to complimenting the user herself for being healthy or fit.


\begin{figure}[h]
\centering
\includegraphics[scale=.2]{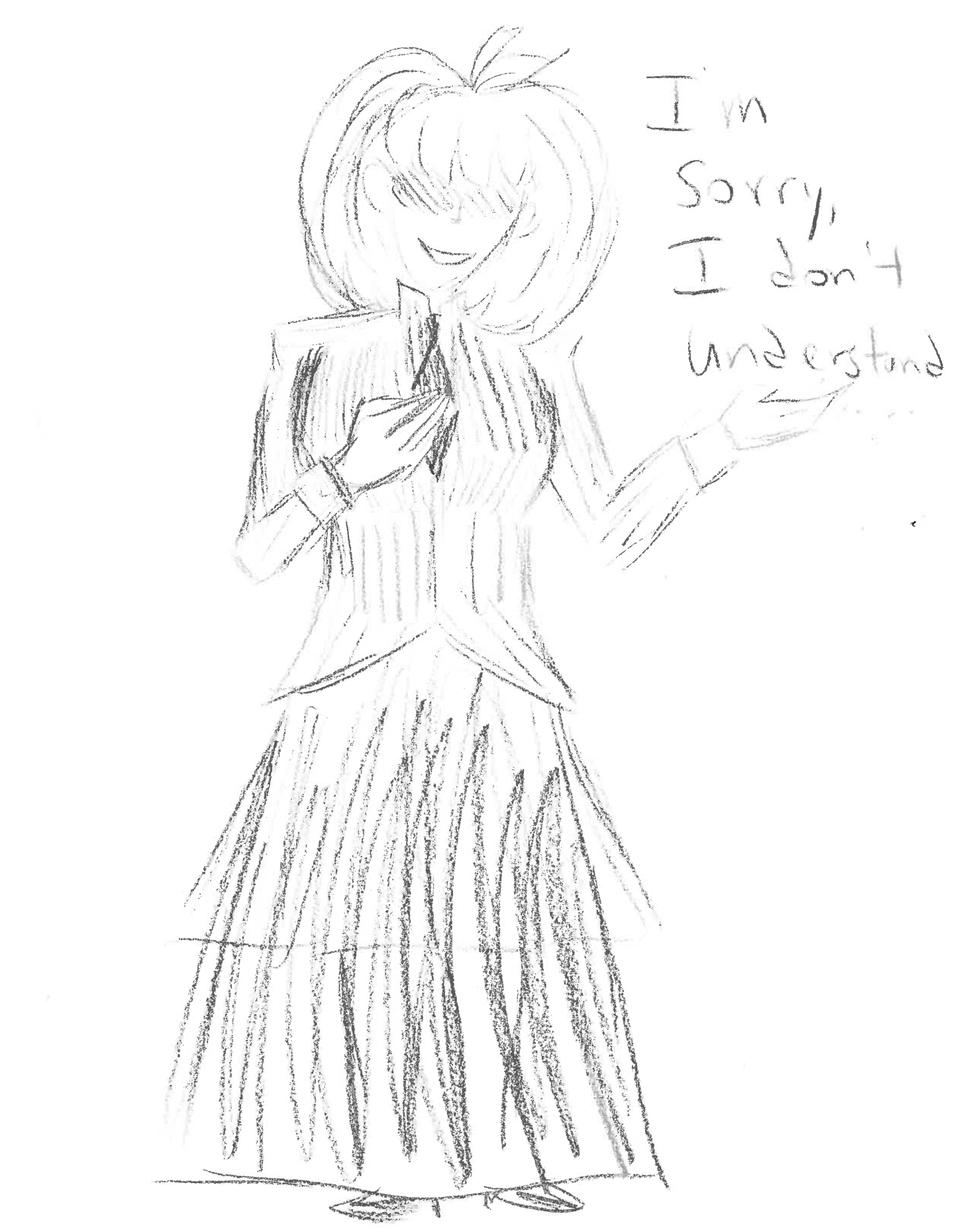}
\caption{\textbf{Some participants were weary of the corporate ties assistants had.} P3 alludes to Apple Inc, by drawing Siri with an apple head. Similar to P16's assistant (Figure \ref{fig:P16}), Siri's eyes are covered to denote the assistant's insincerity and non-human elements \cite{megias2020reading}. Siri's clothing is a gothic business style, alluding to sinister corporate behaviors and data practices.} 
\label{fig:P3} 
\end{figure}

\paragraph{Corporate Inauthenticity} \label{sec:corporate}
 Some participants expressed unease at the idea of their assistant giving affirming messages. When prompted to write affirming messages, P1 said they \textit{``don’t want [their assistant] to feel too close.''} P12 explains how messages that are too positive and personal can be \textit{``very agitating''} and \textit{``offensive.''} She elaborated saying that \textit{``The machine does not know the user. How come they're making these weird, personal comments.''} P7 had similar thoughts about her Google Assistant, expressing that \textit{``because it’s Google and a product, it would feel kind of strange with a corporation to literally be like, `we hear you, but we don’t understand you.'''} A couple participants were cognizant of the corporations behind their voice assistants to the extent that they integrated them in their drawings. For example, P3 drew a feminine figure with an apple-shaped head (referencing Apple's Siri) and the words ``Sorry, I don't understand what you said'' emerging out of a smiling mouth (See Figure \ref{fig:P3}). This sentiment may explain why many of the affirming messages participants wrote were more subtle rather than direct.




\section{Guidelines for Harm-Reducing Conversational Designs}

In the following section, we interpret our findings through the lens of prior empirical studies and theories to provide recommendations for harm reduction and repair. The recommendations given here provide research directions, design considerations, and technical priorities, and thus may be relevant to an array of researchers, designers, and engineers. 

\subsection{Affirm Complementary Identities} While identity affirmations are known to be restorative to individuals facing relevant harms \cite{martens2006combating, wenzel2023can, sherman2006psychology}, designers must be mindful when deciding which part of an individual's identity an affirmation refers to. Affirmations that reference the identity attribute implicated in an instance of miscommunication (which, in this case, is likely to be ethnicity/language), can amplify the harm an individual experiences. This was not only revealed in our study, with participants expressing unease at cultural and ethnic-related affirmations, but it is also corroborated by prior research from psychology \cite{aronson1995dissonance, mccrea2011limitations}. When part of individual's identity is psychologically threatened, reminding her of that identity's positive ideal (via affirmation) may trigger the \textit{self-discrepancy effect}. Through this effect, individuals compare their perceived ``actual'' self to their ideal self. Harms position individuals' ``actual'' self at a deficit, while affirmations raise the standard of their ideal self (e.g., attributes of users' identity that align with their core values or self-related interests and pursuits). This combination can widen the gap between individuals' actual and ideal selves, which results in a sense of self-discrepancy and feelings of unworthiness \cite{aronson1995dissonance}. Affirmations may, in fact, be best effective when they go undetected by the recipient \cite{sherman2009affirmed}.

Thus, it is recommended that affirmations in voice interactions provide reminders of \textit{complementary} positive attributes and values. For participants in our study, these complementary values typically revolved around media and entertainment culture, which can represent significant components of one's personal identity and important bases for self-expression and self-verification \cite{haggard1992identity, loersch2013unraveling}. Understanding what complementary attribute may be appropriate to affirm may be based on data the user offers to the voice assistant provider either through an onboarding experience or inferred through patterns present in users' everyday interactions with their assistant.

\subsection{Be Culturally Sensitive}
When designing affirmations, or communication repair strategies in general, it is important to heed respect to diverse user cultures. Several researchers have put forth recommendations for approaching multi- and cross-culturally user needs within a single corporation or product \cite{marcus2006cross, de2016influences, yiu2013ux, marcus2014cross, lachner2015cross}. One approach that has emerged from this work encourages employing Hofstede's cultural dimensions model \cite{marcus2006cross, hofstede2011dimensionalizing}, which may help practitioners design for differences in individualistic vs. collectivist cultures, high vs. low uncertainty avoidance cultures, long- vs short-term time orientation cultures, among others. Understanding these diverse cultural needs can help address instances of dissatisfaction and aversion, such as what P5 expressed in Section \ref{sec:aware}. 

Voice interactions and conversational AI more broadly should also be culturally sensitive to the extent that it avoids identity \& cultural harms (Section \ref{sec:culture-harm}). Addressing this specific harm can be difficult because it is, to some extent, tied to technical limitations. However, as we and many other researchers continue to reveal the material identity \& cultural harms that emerge from these technologies, we urge that addressing such harms be prioritized in the development pipeline. One profound action item would be to increase the database of proper nouns used in language technologies. The misrecognition of non-Anglo names has been a persistent harm across many language technologies, including voice interfaces, for many years, as documented by our participants and by many others \cite{dyal2021autocorrecting, benjamin2020race}.

\subsection{Redirect Blame}
In addition to affirming complementary identities and being culturally sensitive, we also suggest blame redirection as an impactful recovery strategy. Blame redirection may include an apology given by the assistant and/or an explanation describing the error. To properly assume blame with an explanation, it is imperative that this explanation describes a fault or misalignment of the assistant rather than any fault of the user, be it real or contrived. 


\subsection{Adjust Delivery (Timing \& Frequency)} 
 Importantly, repair messages need to balance harm-reducing content (i.e. apologies and affirmations) with brevity, and furthermore, affirming messages should not be too frequent. Voice assistants are intended to afford convenient, hands-free interactions. Not only is this built into their design \cite{guzman2019voices}, but our participants could clearly articulate this affordance as well. As such, designers should work to negotiate affirming content into concise messages.

 Another important issue of concern is the frequency of affirming messages. If affirmations are as powerful as social psychology research suggests \cite{sherman2006psychology, cohen2014psychology}, should they be employed during every communication breakdown with an assistant? While more research is needed to verify the appropriate rate of affirmation delivery, we currently suggest that affirmations be delivered \emph{intermittently}. Employing affirmations too frequently may make users' suspicious of the harm-reduction intervention. This awareness, whether it be subconscious or conscious, can elicit what Brehm termed ``psychological reactance,'' in which users become attitudinally and behaviorally resistant to the intervention, even if it is to their benefit \cite{steindl2015understanding}. As mentioned earlier, affirmations are most beneficial when they are undetected \cite{sherman2009affirmed}. Future work is needed to determine the frequency that simultaneously maximizes the benefits of affirmations and minimizes the chance of users detecting them.
 
 

\subsection{Guideline Limitations}

Note that the suggestions we offer here primarily target psychology-related harms, that is, any harm that may affect a user's social or cognitive perceptions, attitudes, or feelings. This is inclusive of many of the harms we have identified. However, there are a few harms that require more material interventions, such as physical endangerment. We do not intend to suggest that an affirmation may prevent a user from being harmed by a car crash. However, it is possible that the psychological elements associated with the physical endangerment harm may be partially addressed. For example, our guidelines may help ameliorate road rage, which is associated with both voice navigation frustrations and car crashes, respectively.

\section{Discussion}

\subsection{Overview of Contributions}

Our research offers a novel approaches multiple aspects of the voice assistant user experience. Firstly, we use visual and qualitative methods to understand how multicultural users' frustrations with their voice assistant affect their conceptualizations of their assistant. While prior work has highlighted the various ways in which users may anthropomorphize their assistant, to our knowledge this is the first study that focuses specifically on users that have been \textit{underserved} by voice assistants. Despite our focus on a specific population, there is potential for some of our findings to generalize to mainstream users as well, akin to digital curb cuts \cite{petrick2019curb}. Secondly, we apply a harm framework to the voice assistant user experience, focusing specifically on the experience of a high error rate. To the best of our understanding, this is the first instance that scopes significantly beyond quality-of-service harms when examining a technology that has a high error rate. Thirdly, we combine findings from our participatory research with empirical evidence from psychology to offer psychologically harm-reducing communication repair strategies. Our work merges and subscribes to two growing research agendas (1) harm reduction and (2) communication repair.

Through our qualitative study, we also corroborated and strengthened confidence in insights from prior work, and we offer this as our final contribution. Our findings related to users' code-switching behaviors has been highlighted in prior work on Black voice assistant users \cite{harrington2022s, mengesha2021don}, and our study extends the phenomenon to users of other ethnicities and cultural backgrounds. Furthermore, our work contributes to the small, but important, area of research on non-American voice assistant users \cite{lima2019empirical, choi2023toward, wagner2019human, seymour2021exploring}, as our users confirmed frustrations with Korean- and Indian-based voice assistants. Lastly, our suggestion to redirect blame and employ brevity has been highlighted by other researchers \cite{mahmood2022owning, cuadra2021my, haas2022keep}, and we are excited to support innovation in this direction.




\subsection{Voice Assistants Are Not (Necessarily) Friends}
Prior work has found that user tendencies toward anthropomorphism diminish as users realize the limitations of their voice assistant \cite{cho2019once}. However, little prior work has focused on the anthropomorphic tendencies of users who are notably underserved by voice assistants, such as the population of our study. When explicitly prompted to anthropomorphize their assistant, our participants tended towards depicting secretary-like characters and/or figures who had trouble processing emotional and social cues. These attributes are not novel in-and-of themselves, yet, this finding is still interesting in that it demonstrates how (1) these two types of associations persist across cultures and (2) positive associations, which researchers have documented in other populations (see Section \ref{sec:anthro}), do not necessarily persist across cultures. This result is likely due to the usability issues and harms our participant population faces with voice assistants. 

\subsection{High Error Rates in Voice Assistants are Not Only Quality-of-Service Harms}
Our findings demonstrate that voice assistants encapsulate many harms already documented by previous quality-of-service harm frameworks (i.e. increased labor, service/benefit loss \cite{shelby2022sociotechnical}; identity \& culture harms \cite{dyal2021autocorrecting}). We also found harms that have specifically been documented in prior voice assistant literature (i.e. relational harm \cite{storer2020all} and physical endangerment \cite{bickmore2018patient}). However, we stress here that relational and physical harms as documented in prior work did not relate these harms to the high error rate that marginalized users experience. In the case of Storer et al.'s relational harm, their participants were concerned about a variety of experiences unrelated to accuracy. They were worried that their voice assistant would replace quality interpersonal interactions, were concerned about unauthorized purchases on shared home devices, and were startled by loud intercom uses of assistants. While all of these instances of relational harm should certainly be avoided, none of them are related to speech-recognition error rates. Similarly, the physical health harms Bickmore et al. identified were often due to improper capture, insufficient sources, or lack of medical disclaimers, which are unique from the problem of inaccurate speech recognition related to one's marginalized identity.

\subsection{Limitations \& Future Work}

\subsubsection{User Evaluation} As a next step for communication repair redesign, we suggest a full evaluation of the repair strategies offered in this paper. This would involve crafting communication repair dialogues based on our guidelines, creating comparable dialogues that lack these elements to serve as a control, and running an experiment that evaluate user responses and reactions to the various experimental conditions. While we believe our interventions are promising, as they are guided by both participatory work and prior empirical studies, we advocate for rigorous testing and evaluation before deployment, in line with the harm-reduction agenda.


 \subsubsection{Generalizing Harm Reduction Toward Other Technologies}
In the present study, we specifically studied harm reduction techniques for voice assistants. While the scope of this technology is limited, we believe our interventions have promise in other contexts. For example, in text-based language technologies such as auto-correct and auto-reply, which are also embedded with social biases \cite{dyal2021autocorrecting}, suggestions may be accompanied with an abbreviated blame-redirecting message which emphasizes the limits of the technology's abilities. If a user rejects a suggestion, a message of affirmation may briefly appear to reassure the user of their identity and choice. These techniques may even extend beyond AI-based language technologies and be adapted into error messages for various technological services and goods more broadly. Such interventions deserve further research and evaluation to inform their proper design.

\subsubsection{{Taxonomizing Harms}}
The harms we offered in this paper were specifically related to high speech recognition error rates for multicultural voice assistant users. We chose this focus as we believe multicultural and multilingual users specifically are underserved in the AI voice technology development pipeline, and furthermore we wanted to demonstrate that error rates can do much more than reduce the basic usability of a product. We stand by this framing and believe it is important. Of course, we also acknowledge that there are many other voice assistant harms unrelated to error rate that are also deserving of attention, such as privacy and surveillance violations \cite{zuboff2023age, lee2023deepfakes, woods2018asking} or misinformation concerns \cite{bickmore2018patient}. A comprehensive voice assistant harm documentation has still yet to be made. 

\subsubsection{Intersectionality and Defining ``Multicultural''}
While we recruited participants who identified as multicultural or multilingual, we would like to highlight how these participants had other intersecting identities. For example, participants may have identified as immigrants, expatriates, non-American citizens, non-native English speakers, etc. These various identities, while sometimes related, are distinct from the labels of multicultural and multilingual. The results we present in our study may be connected to any of our participants' unique identity characteristics, however due to the nature of intersectional identities, it is difficult to pinpoint the exact connections. Future work may consider capturing these granularities.

\subsubsection{Sample Population}
Our qualifying participant recruitment criteria was determined by finding commonalities between the research team's own identities, honoring feminist standpoint theory \cite{bardzell2011towards}, and users who have had demonstrated inequities with voice technologies. While we aimed to recruit a diverse set of participants, our participants were still limited in a few ways. Our participants largely self-identified with Indian and American identities. A few identified with other parts of the Asian diaspora, and beyond this, we had four participants identify as queer, Ashkenazi Jewish, Black, and Mexican, respectively. Future work should consider how the harms we identified may generalize to those who were in the minority of our sample or absent from our sample entirely. 

Regarding age, with the exception of P12, all of our participants were young adults (18-29 years old). Furthermore, most of our participants were students. Future studies may aim to understand how the presented harm reduction principles may be adapted to older adults, children, and non-students. 

Only two of our participants identified as non-binary and transgender. Rincón et al. \cite{rincon2021speaking}, has already examined needs and explored the tensions of non-binary and transgender voice assistant users, future research may then focus on designing specifically for communication repair for trans and non-binary folk. Naturally, there are a myriad of other identities that we have not touched upon in this work, including but not limited to, indigenous users, neurodivergent users, users without higher education, et cetera. Future work should respect and explore these user groups in more depth.

\section{Conclusion}
The present study deployed an interview / design workshop study with 16 multicultural voice assistant users. Our contributions include an understanding of multicultural users conceptualizations of their voice assistant (knowledgeable, incompetant, unemotional, lacking social skills), a harm framework that focuses on the unique case of high speech-recognition error rate, and evidence-informed design guidelines for harm-reducing communication repair (intermittent identity affirmations, cultural sensitivity, blame redirection, brevity). Our work offers insights to research agendas related to harm reduction, communication repair, and multicultural/multilingual/marginalized users.


\begin{acks}
 We would like to thank Jiwon Lee, Alicia Ng, and Victoria Santiago who helped select the workshop activities. We would also like to thank research assistants Lukas Chen and Yvonne Huang who helped deploy our early workshops. Lastly, thanks to Laura Dabbish for her suggestions pertaining to this work. This research was supported by National Science Foundation under Grant \#2040926.

\end{acks}

\bibliographystyle{ACM-Reference-Format}
\bibliography{references.bib}

\section{Appendix}
\begin{appendix}
\appendix

\textbf{Study protocol and Facilitator script}

\textbf{Directed Storytelling (10 minutes):}  In this activity, we start off by asking participants to \textit{“Tell a story about a time you felt frustrated interacting with a voice assistant. Please reference the following elements when telling your story, and be as specific as possible.”} 4 Elements to reference written on a white board: Participants are encouraged to be as specific as possible in terms of 1.Type/name of voice assistant, 2.Context of interaction 3.Any recollection on the choice of words they used 4.Responses given by voice assistant. 
Potential guiding questions for the facilitator:
\begin{itemize}
    \item How often do you use that voice assistant now?
    \item What do you use your voice assistant for?
    \item Why do you think you had such an experience?
    \item Do you think your experience with the voice assistant is unique?
    \item If there is one thing you could change about your voice assistant, what would it be?
\end{itemize}

\textbf{Drawing a representation of your voice assistant (5 minutes):} Drawing materials will be provided to the participant, and they will draw a personification/character of their voice assistant on a piece of paper. \textit{“Using the drawing materials provided to you, please draw a personified version of your voice assistant. We encourage you to be creative with this activity. Consider: If your voice assistant was a person, what kind of person would they be? What would they look like? What kind of personality would they have? What types of voice assistantlues would they hold? And remember, there is no wrong way to draw!”} The goal of the activity is to allow participants to develop a visual representation of their voice assistant based on their prior experiences. This activity leads up to the next activity where participants will direct their thoughts and feelings to their voice assistant via a Love/Break up letter.

\textbf{Love/Break up Letter (10 minutes):} Pen and paper will be distributed to each participant. Participants will be asked to write a love/break-up letter to the voice assistant representation they have rendered in the previous activity. \textit{“Take a moment to reflect on your experiences with your voice assistant, or a voice assistant you’ve used in the past. Based on your feelings and experiences, write either a love letter or a breakup letter to the voice assistant. Imagine if they were a person; what would you say to them?”} Each participant will be given 5 minutes to write the letter. After 5 minutes are up, users have a choice to read their letter out loud and share their sentiments. After the participants have finished the letter, the letters will be collected. 

\textbf{Creating a Cultured Virtual Assistant (15 minutes):} In this activity, participants will be asked to create the ideal response that they believe a voice assistant should say. Participants will make a set of 6 cards. Some of the cards will be freeform. The other set of cards will be responses that support the participants’ own identity. We will discuss the participant's responses and collect the response cards after the activity is completed. \textit{“Now that you have expressed your emotions towards your virtual assistant, let’s think of how we might improve your assistant. Think– When your voice assistant misunderstands you or produces an error, what would you like it to say to you? More specifically, how might your voice assistant best respond to you or people from a similar background? In front of you are a set of cards. On each card, please write out your ideal voice assistant error response.} [REFERENCE BOARD WITH DIRECTIONS] \textit{Some of the cards will be freeform– you can write whatever you want! The other cards will employ identity affirmations. Identity affirmations help individuals foster positive feelings and a sense of belonging to their social group. For example, instead of just reporting an error or saying ‘sorry,’ your assistant might give a message reinforcing your positive attributes, it might let you know that they think you’re cool, or acknowledge your broader culture. Oftentimes, affirmations are associated with encouragement, compliments, and warmth. During the next 5-10 minutes, please write at least 2 freeform responses and 4 identity affirming responses that you would like to hear from your assistant. If you can’t fill up all of the cards, don’t worry, but see it as a goal.”}

\textbf{Debrief (5 minutes):} Thank participants for their participation. Allow participants to reflect on their experiences. What insights have they gained? Potential guiding questions for the facilitator:

\begin{itemize}
    \item What is the biggest takeaway you got from this workshop?
    \item Are there alternative methods or responses, besides identity affirmations, that you would like to see implemented in voice assistants?
    \item Is there anything else you would like to change about voice assistants? 
\end{itemize}

\begin{table*}[h!] 
\caption{Additional self-reported demographics of all participants. Languages are listed in order of dominance. For Employment status: (S=Student, E=Employed, U=Unknown). Among the students, there was a mix of disciplines and degree programs. Both undergraduate and graduate students were represented. For Gender: (M=Man, W=Woman, N=Non-binary). Transgender is binarily coded: (N=No, Y=Yes).}
\begin{tabular}{|c|l|l|c|c|c|}
\hline
\textbf{ID} & \textbf{Employment}                        &               \textbf{Age}               & \textbf{Gender} & \textbf{Transgender}               \\ \hline
P1          & E                                   & 18-29                               & M               & N                                        \\ \hline
P2          & S             & 18-29                                            & M               & N                                      \\ \hline
P3          & S                         & 18-29                       & N               & Y                                         \\ \hline
P4          & S              & 18-29 & W               & N                                         \\ \hline
P5          & S        & 18-29               & W               & N              \\ \hline
P6          & S                            & 18-29                                            & M               & N                                       \\ \hline
P7          & S                  & 18-29                       & N               & Y                                         \\ \hline
P8          & S                            & 18-29                                            & M               & N                                          \\ \hline
P9          & S                    & 18-29                       & M               & U                                         \\ \hline
P10          & S & 18-29                                           & W               & N                     \\ \hline
P11         & S                     & 18-29                                            & M               & N                     \\ \hline
P12         & U                         & 40-49                              & W               & N                    \\ \hline
P13         & S          & 18-29                                            & M               & N                     \\ \hline
P14         & S                          & 18-29                              & M               & N                    \\ \hline
P15         & S                 & 18-29                    & M               & N                    \\ \hline
P16         & S               & 18-29                  & W               & N                   \\ \hline
\end{tabular}
\label{table:demographics-2}
\end{table*}

\end{appendix}

\end{document}